\documentclass[aps,pra,showpacs,amsmath,amssymb,amsfonts,tightenlines,twocolumn,lengthcheck,superscriptaddress]{revtex4-1}
\usepackage{amsmath}
\usepackage{amsfonts}
\usepackage{graphicx}
\usepackage{epsfig}
\usepackage{color}
\usepackage{txfonts}
\usepackage[colorlinks,citecolor=blue]{hyperref}

\begin{document}

\title{Controlling atom-photon bound states in a coupled resonator array with a two-level quantum emitter}
\author{Zelin \surname{Lu}}
\affiliation{Synergetic Innovation Center for Quantum Effects and Applications, Key Laboratory for Matter Microstructure and Function of Hunan Province, Key Laboratory of Low-Dimensional Quantum Structures and Quantum Control of Ministry of Education, School of Physics and Electronics, Hunan Normal University, Changsha 410081, China}
\author{Jing \surname{Li} }
\affiliation{Synergetic Innovation Center for Quantum Effects and Applications, Key Laboratory for Matter Microstructure and Function of Hunan Province, Key Laboratory of Low-Dimensional Quantum Structures and Quantum Control of Ministry of Education, School of Physics and Electronics, Hunan Normal University, Changsha 410081, China}
\author{Jing  \surname{Lu} }
\affiliation{Synergetic Innovation Center for Quantum Effects and Applications, Key Laboratory for Matter Microstructure and Function of Hunan Province, Key Laboratory of Low-Dimensional Quantum Structures and Quantum Control of Ministry of Education, School of Physics and Electronics, Hunan Normal University, Changsha 410081, China}
\author{Lan \surname{Zhou}}
\thanks{Corresponding author}
\email{zhoulan@hunnu.edu.cn}
\affiliation{Synergetic Innovation Center for Quantum Effects and Applications, Key Laboratory for Matter Microstructure and Function of Hunan Province, Key Laboratory of Low-Dimensional Quantum Structures and Quantum Control of Ministry of Education, School of Physics and Electronics, Hunan Normal University, Changsha 410081, China}

\begin{abstract}
We consider a one-dimensional (1D) coupled-resonator array (CRA), where a two-level quantum
emitter (2LE) is electric-dipole coupled to the modes of two adjacent resonators. We investigate
the energy spectrum, the photon probability distribution of the bound states and the emission
process of the 2LE into the CRA vacuum. A quantum phase transition is found which is characterized
by the change of the number of the out-of-band discrete levels. The condition for this change is also presented. The photon wave functions of bound states are found to be asymmetry around the position of the 2LE when the coupling strengths between the 2LE and the resonator are not equal, and they have the same preferred directions which are primary determined by the larger one among the coupling strengths. The presence of the atom-photon bound states is manifested in the form of
a stationary oscillation or a non-vanishing constant in the long enough time.
\end{abstract}


\date{\today}
\maketitle
\section{Introduction}
Light-matter interaction which are the essential physics of many phenomena,  have long been of great interest because of their potential applications in quantum information processing.  Natural and artificial atoms are generally used to encode information due to their considerable time scales to coherently storing information in this process. Photons in comparison with other information carriers are very effective
to transfer information due to their high transport speed for distributed quantum information. The ability
to enhance the interaction between atoms and photons lies at the heart of the achievement of state transfer.

A high-finesse resonator insulates photons from the environment and stores photons for a long time before
they are dissipated, which leads to a strong coupling between atoms and photons. Recently, systems of coupled
resonators have received much attention\cite%
{BoydPRA69,KimPRA78,XPRL102,Zhou08,zhouPRA78(08),gongPRA78(08),Zhou09,104(10)023602,zhouPRA85(12),Zhou13,lujing89,wangPRA89,xuPRA17,MahPRL123,LJinPRR4}%
due to their available control of individual resonators and coupling of light
at slow group velocities. Coupled-resonator arrays (CRAs) typically are an
arrangement of low-loss resonators with nearest-neighbor coupling, where
photons can hop across adjacent resonators. Resonators in such array are
handled as individual sites, so a one dimensional (1D) CRA is usually described by the tight-bind model~%
\cite{Zhou08,zhouPRA78(08),gongPRA78(08),Zhou09,104(10)023602,zhouPRA85(12),Zhou13,lujing89,wangPRA89,xuPRA17,MahPRL123,LJinPRR4,PRX12BS,PRA102BS,PRA108 053717,PRA108 013704}%
. Besides operations on photonic qubits traveling in CRA are exploited, such as switch~\cite{Zhou08,Zhou09}, trapping~\cite{zhouPRA78(08),gongPRA78(08)}, router~\cite{Zhou13,lujing89}, frequency converter~\cite{wangPRA89}. Moreover, a richer spectrum
structure is found ---- the finite width of the band, out-of-band discrete
levels where photonic distribution is localized around the single quantum emitter~\cite{Zhou09,PRX12BS,WangPRA101,PalmaPRA89,PRA100(19)063806,PRA96(17)023831}.
In this work, we study the system of a point-like two-level emitter (2LE) simultaneously coupled to
the EM field modes of two adjacent resonators in a 1D CRA. We find that the photonic wave function of the bound states is symmetrically
or unsymmetrically distributed around the two resonators coupled to the 2LE depending
on whether the two coupling strengths are not equal, and one can
control the number of out-of-band bound states by adjusting the coupling strengths and
the transition frequency of the 2LE, which reveals the occurrence of a quantum phase
transition. As it seems hard to excite out-of-band discrete levels through photon scattering
in the single excitation subspace, we study the occupation of the 2LE's excited state
with the CRA initially in the vacuum state, the presence of the bound states is manifested
in the form of residual oscillatory or non-zero constant behavior of the excited state
population at long times.

\begin{figure}[tbp]
\includegraphics[width=0.4\textwidth]{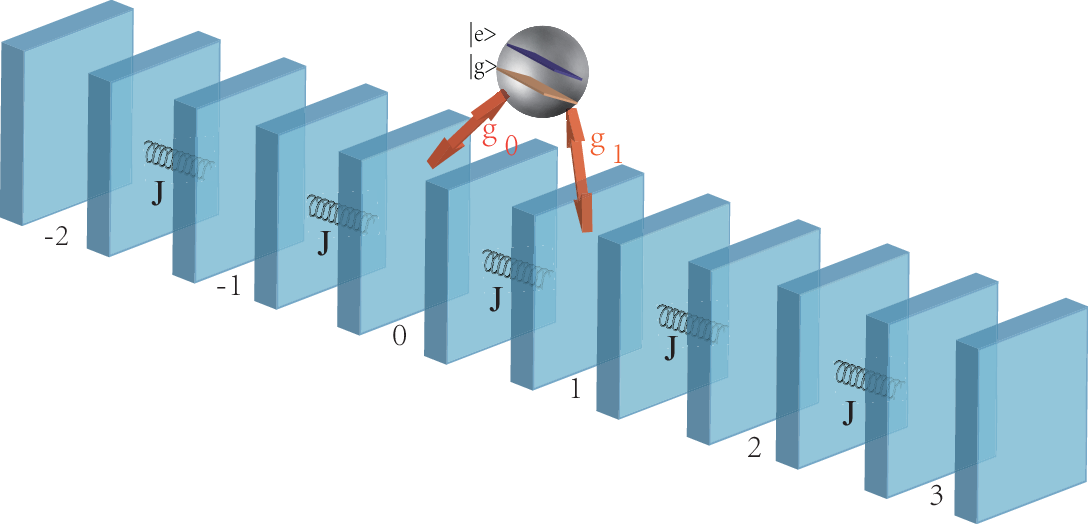}
\caption{(color online) Sketch of the setup: a coupled resonator array (the blue rectangular plates)  coupled to a 2LE.}
\label{fig1}
\end{figure}

As shown in Fig.~\ref{fig1}, the system under study consists of a 1D array of
identical, single-mode resonators and a 2LE with an excited state $|e\rangle$
and ground state $|g\rangle$. Nearest-neighbor resonators are assumed to be close
enough to ensure the mutual transfer of photons via evanescent fields. Each resonator
is modeled as a single harmonic oscillator mode of frequency $\omega_{c}$. By treating the array as a tight-binding model with a hopping
strength $J$, the free Hamiltonian of the system reads
\begin{equation}
\hat{H}_{0}=\Omega \hat{\sigma}_{+}\hat{\sigma}_{-}+\sum_{j
}\left[ \omega _{c}\hat{a}_{j}^{\dag }\hat{a}_{j}-J\left( \hat{a}%
_{j+1}^{\dag }\hat{a}_{j}+\mathrm{H.c.}\right) \right] ,
\label{1-01}
\end{equation}
where $\hat{\sigma}_{-} (\hat{\sigma}_{+}) $ is the lowing (rising) operator, $\hat{a}_{j} (\hat{a}_{j}^{\dag })$  annihilates (creates) a photon at the $j$th resonator. The Hamiltonian for the 2LE with  Bohr frequency $\Omega$ coupled to the modes of two adjacent resonators with corresponding strengths $g_{0}$ and $g_{1}$ reads
\begin{equation}
\hat{H}_{1}=\left( g_{0}\hat{a}_{0}^{\dag }+g_{1}\hat{a}_{1}^{\dag }\right)\hat{\sigma}_{-}+\mathrm{%
H.c.},
\label{1-02}
\end{equation}
where $g_{0}$ and $g_{1}$ are set to be real. The system Hamiltonian $H=H_{0}+H_{1}$ entails conservation of the total number
of excitations. The eigenstates in the single excitation subspace have the form
\begin{equation}
\left\vert \Psi \right\rangle=\sum_{j=-\infty }^{\infty }\alpha _{j}\hat{a}%
_{j}^{\dagger }\left\vert \emptyset \right\rangle +u_{e}\hat{\sigma}%
_{+}\left\vert \emptyset \right\rangle ,
\label{1-03}
\end{equation}%
where $\left\vert \emptyset \right\rangle =\left\vert 0g\right\rangle $ is the state
with the 2LE in ground states and field in the vacuum state, and $\alpha _{j}$, $u_{e}$
are the corresponding amplitudes. The time-independent Schr\"{o}dinger equation yields
the equations for the amplitudes as
\begin{subequations}
\label{1-05}
\begin{eqnarray}
E\alpha _{j}&=&\omega _{c}\alpha _{j}-J\left( \alpha _{j+1}+\alpha
_{j-1}\right) +g_{0}u_{e}\delta _{j0}+g_{1}u_{e}\delta _{j1}, \\
Eu_{e}&=&g_{0}\alpha _{0}+g_{1}\alpha _{1}+\Omega u_{e}.
\end{eqnarray}%
\end{subequations}

\section{\label{Sec:2} The single-excitation spectrum. }
\begin{figure}[tbp]
\includegraphics[width=0.4\textwidth]{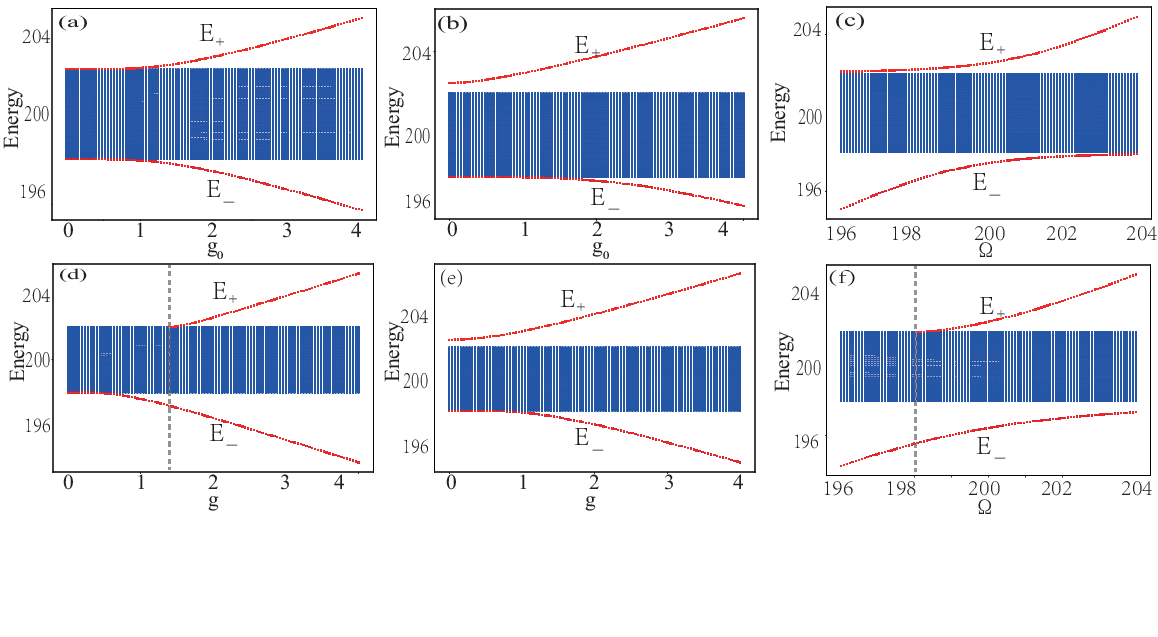}
\caption{(Color online) The single-excitation spectrum versus coupling strength $g_{0}$
with $g_{1}=0$ in (a, b), $g_{1}=g_{0}=g$ in (d, e) and the atomic transition frequency
$\Omega$ in (c, f) for $g_0=g_{1}=2$. The other parameters are set as follow: $\omega_c =200$,
$\Omega =200$ for (a, d) and $\Omega =202.5$ for (b, e). All parameters are in units of the hop
strength $J$, so do the following figures.}
\label{fig2}
\end{figure}

Eq.~(\ref{1-05}) has two types of atom-photon dressed states: the single-photon
spatially-localized states called bound states and the following scattering eigenstates
spatially extended over the whole waveguide
\begin{eqnarray}
\label{2-01}
\left\vert \psi _{k}\right\rangle&=&\sum_{j}\left[ \left(
e^{ikj}+r_{k}e^{-ikj}\right) \theta \left( -1-j\right) +t_{k}e^{ikj}\theta
\left( j\right) \right] \hat{a}_{j}^{\dagger }\left\vert \emptyset
\right\rangle \notag \\
&&+\frac{g_{0}\left( r_{k}+1\right) +g_{1}t_{k}\exp \left( ik\right) }{%
\omega _{k}-\Omega }\hat{\sigma}_{+}\left\vert \emptyset \right\rangle,
\end{eqnarray}
where $\theta(x)=0$ for $x<0$ and $\theta(x)=1$ for $x\geq 0$, the transmission
amplitude $t_k$ and the reflection amplitude $r_k$ are given
in Eq.(9) and (10) in Ref.~\cite{zhouPRA85(12)}. So the spectrum of the system consists
of a continuum and discrete levels. The energy of the continuum form an energy band, whose
relation with the wave number $k\in[-\pi,\pi]$ within the first Brillouin zone is given by
$\omega _{k}=\omega _{c}- 2J\cos k$.
In Fig.~\ref{fig2}, we plot the single-excitation spectrum as a function of the coupling
strengths and the atomic transition frequency $\Omega$. Here, two models are considered. The
upper panels (a-c) in Fig.~\ref{fig2} correspond to the system with $g_1=0$, i.e., the CRA
is locally coupled to a 2LE~\cite{Zhou08}. It can be observed that two bound-state energies
$E_{\pm}$ always be presented no matter what values $g_0$ and $\Omega$
take, and energies $E_{\pm}$ move away from the band as $g_0$ increases. When $\Omega=\omega_c$,
two bound-state energies are symmetrically located around the band, as $\Omega$ increases
(decreases) from $\omega_c$, that the upper (lower) bound-state energy move away from the
band more rapidly than the lower (upper) bound-state energy as $g_0$ increases. So we cover
the results in Ref.~\cite{PRA96(17)023831}. The lower panels (d-f) in Fig.~\ref{fig2}
correspond to the system with $g_0=g_1=g$. When $\Omega>\omega_c+2J$ (see Fig.~\ref{fig2}e),
$E_{\pm}$ always exists no matter what values $g$ take. When $\Omega\leq\omega_c+2J$, the bound
state with energy $E_-$ below the band can always be observed. The appearance of the bound
state with energy $E_+$ above the band is dependent on parameters $g$ and $\Omega$, so we
can control the number of bound states.

To understand the appearance and disappearance of the upper bound-state, we introduce
the Fourier transform $\alpha_{j}=\frac{1}{\sqrt{2\pi}}\sum_{k} \beta_{k}e^{ikj}$ to Eq.~(\ref{1-05}a)
and express $\beta_{k}$ with $u_{e}$, then substitute the expression into Eq.~(\ref{1-05}b).
After some algebra, we obtain the energy equation for bound states
\begin{eqnarray}
\label{2-03}
&&\left( E_{\pm}-\Omega -\frac{g_{0}g_{1}}{J}\right) \left( E_{\pm}-\omega_c \right) \sqrt{%
1-\left( \frac{2J}{E_{\pm}-\omega_c}\right) ^{2}}\   \notag \\
&&=g_{0}^{2}+g_{1}^{2}-\frac{g_{0}g_{1}}{J}\left( E_{\pm}-\omega_c\right),
\end{eqnarray}
with energy $|E_{\pm}-\omega _{c}|> 2J$, which indicates that the energies of the bound states
are outside of the band $[\omega _{c}-2J, \omega _{c}+2J]$. Let us denote the left-hand side
of Eq.~(\ref{2-03}) by function $f(E_{\pm})$, and the right-hand side of Eq.~(\ref{2-03}) by $y(E_{\pm})$.
The points where the curves of function $f(E_{\pm})$ and $y(E_{\pm})$ meet determine the number of the
bound states. When $g_1=0$, there are always two points of intersection since function $y(E_{\pm})$
is a positive constant. When $g_1\neq0$, the root $E_y=\omega_c+J(g_{0}^{2}+g_{1}^{2})/(g_{0}g_{1})$
of $y(E_{\pm})=0$ is independent of $\Omega$ and has its lower bound $\omega_c+2J$. The lower bound
of $E_y$ is achieved only when $g_{0}=g_{1}=g$, which indicates that two bound states can be
always found as long as $g_0\neq g_1$. The root $E_f$ of $f(E_{\pm})=0$ is completely determined
by the coupling strengths and the atomic transition frequency, i.e. $E_f=\Omega+g_{0}g_{1}/J$.
When $\Omega+g^{2}/J \leq \omega_c+2J$, the upper bound state disappears. The change
in the energy spectrum indicates that there is a quantum phase transition~\cite{BookQPT}.


\section{\label{Sec:4} Single-photon bound state and its asymmetry  }

\begin{figure}[tbp]
\includegraphics[width=8 cm,clip]{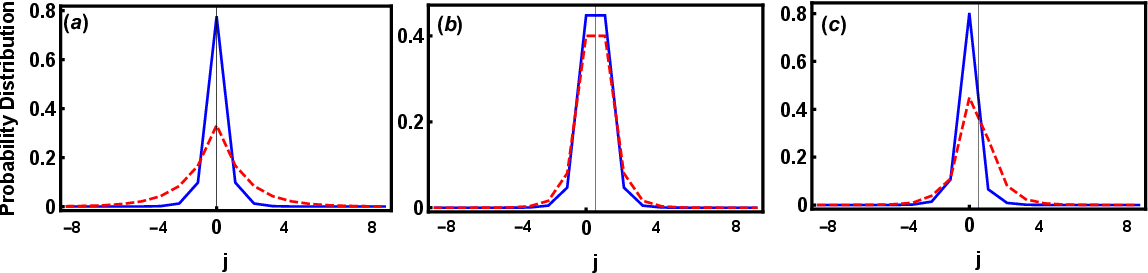}
\caption{(color online) The photonic wave functions of the bound state $\left\vert
\psi _{+}\right\rangle$ (solid lines) and $\left\vert \psi_{-}\right\rangle $
(dashed lines) when $\omega _{c}=200 $, $\Omega=201$. (a) $g_{0}=0.7$, $g_{1}=0$
, (b) $g_{0}=g_{1}=1.7$, (c) $g_{0}=1.7, g_{1}=1$. The black line represents the
symmetrical line of the photonic wave functions.}
\label{fig3}
\end{figure}

\begin{figure}[tbp]
\includegraphics[width=8 cm,clip]{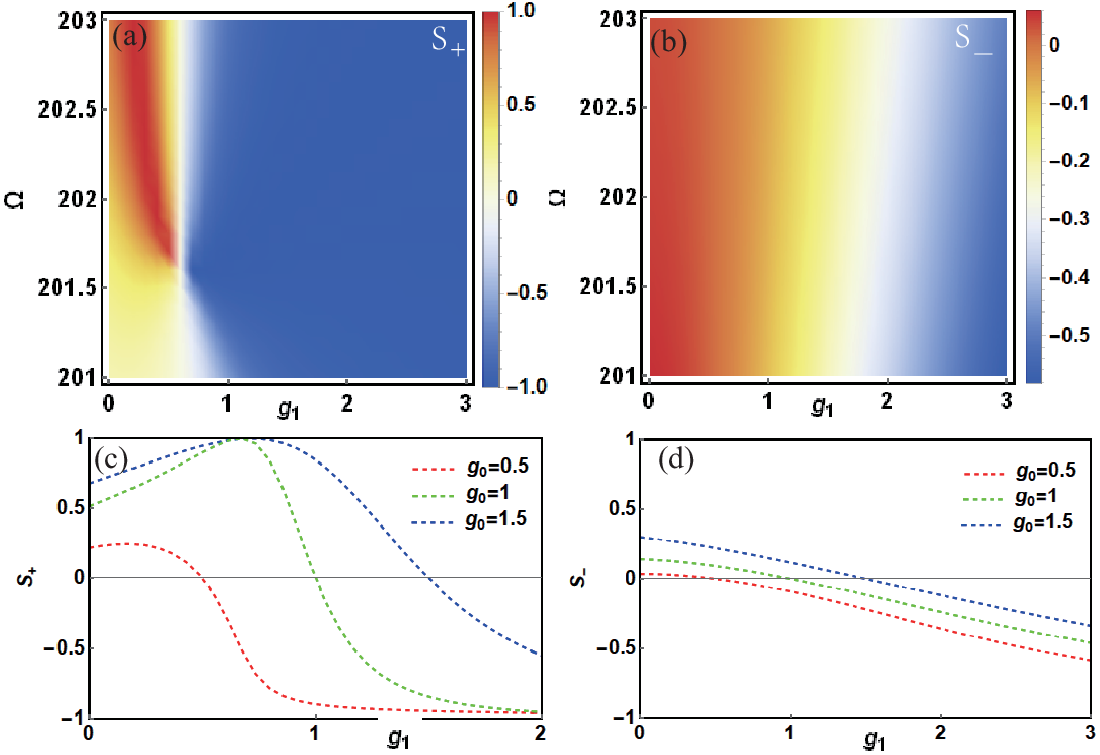}
\caption{(color online)  Tuning the chirality of the bound states when $\omega _{c}=200 $. (a, b) $g_{0} =0.6$, (c, d) $\Omega=201.5$.}
\label{fig4}
\end{figure}

The scattering states in Eq.~(\ref{2-01}) are infinite in spatial extent, but single-photon bound
states
\begin{eqnarray}
\label{3-01}
\left\vert \psi _{\pm }\right\rangle&=&N_{\pm }[\sum_{j=-\infty }^{\infty
}\left( \mp 1\right) ^{j}A_{\pm }^{\theta \left( j\right) }e^{-\kappa_{\pm }
\left\vert j\right\vert }\hat{a}_{j}^{\dagger }  \notag \\
&&+\frac{J\left( g_{0}\mp g_{1}e^{-\kappa _{\pm }}\right) }{J\left( E_{\pm
}-\Omega \right) \mp g_{1}^{2}e^{-\kappa _{\pm }}}\hat{\sigma}%
_{+}]\left\vert \emptyset \right\rangle
\end{eqnarray}%
have wave functions spatially localized around the 2LE's position~\cite{Zhou09,PalmaPRA89,PRA100(19)063806,PRA96(17)023831},
where the factor
\begin{equation}
\label{3-02}
N_{\pm }=\left[ \frac{1+A_{\pm }^{2}e^{-2\kappa _{\pm }}}{1-e^{-2\kappa
_{\pm }}}+\left( \frac{J\left( g_{0}\mp g_{1}e^{-\kappa _{\pm }}\right) }{%
J\left( E_{\pm }-\Omega \right) \mp g_{1}^{2}e^{-\kappa _{\pm }}}\right) ^{2}%
\right] ^{-\frac{1}{2}}
\end{equation}
is a normalization constant with
\begin{equation}
\label{3-03}
A_{\pm }=\frac{J\left(E_{\pm }-\Omega \right) -g_{1}g_{0}}{J\left( E_{\pm }-\Omega \right) \mp
g_{1}^{2}e^{-\kappa _{\pm }}},
\end{equation}
and $A^2_{\pm }e^{-2\kappa _{\pm }}=1$ when $g_{1}=g_{0}$. The parameter $\kappa _{\pm }$ characterizes
the inverse of the localization length and is related to the bound-state energies $E_{\pm }$ by $E_{\pm }=\omega _{c}\pm 2J\cosh \kappa _{\pm }.$
In Fig.~\ref{fig3}, we have plotted the photonic wave functions of state $\left\vert\psi _{+}\right\rangle$
(blue solid line) with energy above the band and state $\left\vert\psi _{-}\right\rangle$ (red dashed line)
with energy below the band by numerical calculation. The photon amplitude decays exponentially away from
the resonators coupled to the 2LE. The wave functions are symmetrically localized round the resonators coupled
to the 2LE when one of the coupling strength vanishes, the axis of the symmetry is indicated by the black
line in Fig.~\ref{fig3}(a). The half of the wave functions at the left side of the black line is also the
reflection of the other half at the right side of the black line in Fig.~\ref{fig3}(b), however, the graph
in Fig.~\ref{fig3}(c) is asymmetrical. We introduce the chirality~\cite{NoriPRL126}
\begin{equation}
\label{3-05}
S=\frac{S_{L}-S_{R}}{S_{L}+S_{R}},
\end{equation}%
to characterize the degree of asymmetry of the bound states. In our system, the vertical line that splits
the graph into two symmetrical mirror images when $g_0=g_1=g$ is located at $j=1/2$, so we define
\begin{equation}
\label{3-06}
S_{L} =\sum_{j=-\infty }^{0}\left\vert \alpha _{j}\right\vert ^{2},
S_{R} =\sum_{j=1}^{+\infty }\left\vert \alpha _{j}\right\vert ^{2}.
\end{equation}
Here, $S>0$ ($S<0$) indicates that the chiral prefers the left (right) direction, and $S\rightarrow 1$
($S\rightarrow -1$) corresponds to perfect left (right) chirality. Since there are two bound states, we
use $S_{\pm}$ to denote the charity of the bound states with energy $E_\pm$. Substituting the wave functions
of Eq.(\ref{3-01}) into Eq.(\ref{3-05}), we obtain
\begin{equation}
\label{3-07}
S_{\pm }=\frac{1-A_{\pm }^{2}e^{-2\kappa_{\pm } }}{1+A_{\pm }^{2}e^{-2\kappa_{\pm } }}.
\end{equation}
To know whether all bound states have the same preferred direction or not, we plot the chirality
of the bound states as a function of the parameters in Fig.~\ref{fig4}.
We emphasize that the asymmetry of wave packet is properly measured by the chirality in Eq.~(\ref{3-06})
when $g_0\neq 0$ and $g_1\neq 0$, however, it is not good for the case with $g_0\neq 0$ and $g_1=0$,
which corresponds to the local coupling. The local coupling only gives rise to symmetrical-distributed
wave packet around the resonator coupled to the 2LE. To properly measure the asymmetry of wave packet
with local coupling, one needs to rewrite Eq.~(\ref{3-06}) by putting the symmetry axe located at the
resonator coupled to the 2LE. The non-vanishing chirality in Fig.~\ref{fig4} for $g_0\neq 0$ and $g_1=0$
is caused by the symmetry axe located at $j=1/2$.

It can be found from Fig.~\ref{fig4} that all bound states have the same preferred direction, the larger
one among the coupling strengths determines the preferred direction of the asymmetry, and bound states are
symmetrical distributed around $j=1/2$ when $g_0=g_1$. Thus, the asymmetry of bound states stems from
the nonlocal coupling of the quantum emitter to two resonators. Although the chirality changes its sign
smoothly as the coupling strength increases, the perfect right (left) chirality can be achieved only
for the upper bound state, and only the chirality of the upper bound state can tune by adjusting the
transition frequency $\Omega$ for given coupling strengths.

The behavior of the 2LE's excitation.
We now consider the dynamics of the system with the 2LE initially prepared
in the excited state $\left\vert \Psi \left(0\right) \right\rangle =\hat{\sigma}_{+}\left\vert \emptyset\right\rangle.$
The time-dependent state for the system of a single excitation shared by the CRA and the 2LE
can be written in terms of all the bound and scattering eigenstates as
\begin{figure}[tbp]
\includegraphics[width=8 cm,clip]{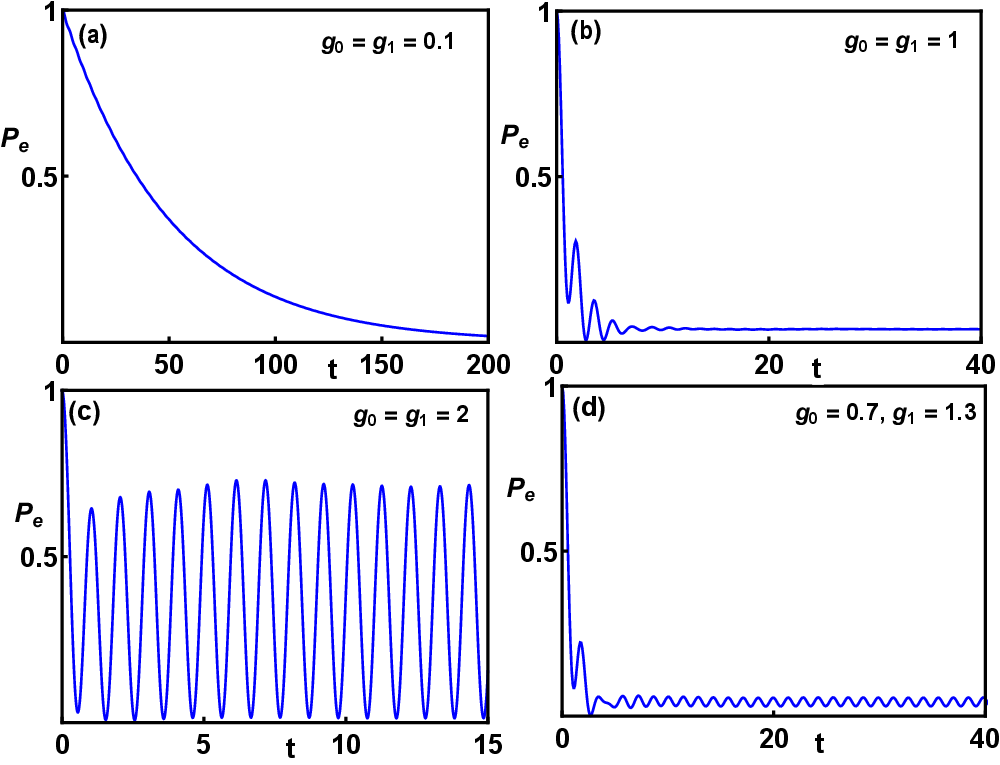}
\caption{(color online)  The probability to find the system in its initial state against time for
different values of the coupling strengths: (a) $g_{0}=g_{1}=0.1$, (b) $g_{0}=g_{1}=1$, (c) $g_{0}=g_{1}=2$,
(d) $g_{0}=0.7,g_{1}=1.3$. Here, $\omega _{c}=\Omega=200$.}
\label{fig5}
\end{figure}
\begin{eqnarray}
\label{4-02}
\left\vert \Psi \left( t\right) \right\rangle &=&\int_{-\pi }^{\pi }\frac{dk%
}{2\pi }c_{k}e^{-i\omega _{k}t}\left\vert \psi _{k}\right\rangle
+c_{+}e^{-iE_{+}t}\left\vert \psi _{+}\right\rangle  \notag \\
&&+c_{-}e^{-iE_{-}t}\left\vert \psi _{-}\right\rangle,
\end{eqnarray}
with
\begin{equation}
c_{k}=\frac{g_{0}\left( r_{k}^{\ast }+1\right) +g_{1}t_{k}^{\ast }e^{-ik}%
}{\omega _{k}-\Omega },  c_{\pm }=\frac{N_{\pm }J\left( g_{0}\mp g_{1}e^{-\kappa _{\pm }}\right) }{%
J\left( E_{\pm }-\Omega \right) \mp g_{1}^{2}e^{-\kappa _{\pm }}}.
\end{equation}

\begin{figure}[tbp]
\includegraphics[width=8 cm,clip]{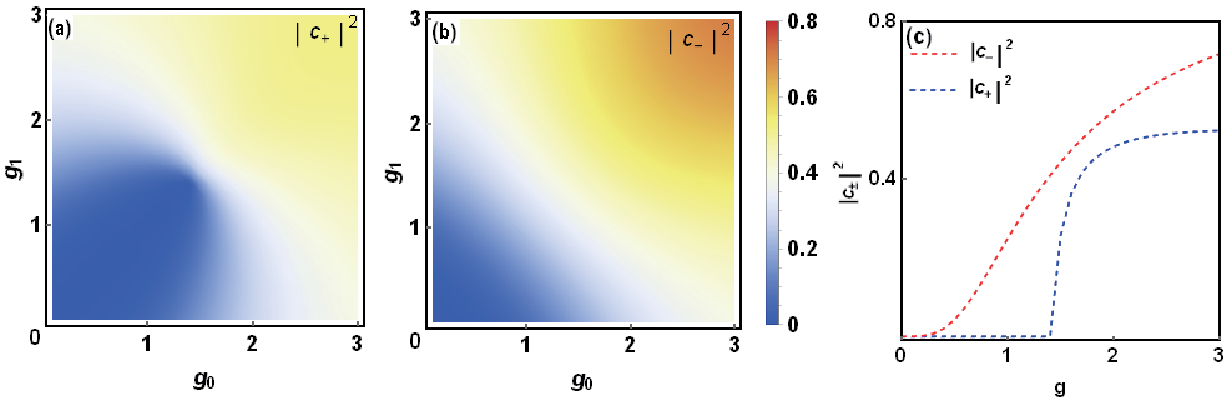}
\caption{(color online)  The probability of bound state versus coupling strengths $g_{0}$ and $g_{1}$ when
$\omega _{c}=\Omega=200$. We set $g_{0}=g_{1}=g$ in (c).}
\label{fig6}
\end{figure}

The light emits from the 2LE would be carried away by a continuum of propagating states. But the
discrete levels make the 2LE in general unable to eventually release its entire amount of initial
excitation to the field, so there still some probability to find the system still in the 2LE at long
enough times. In Fig.~\ref{fig5}, we plot the time evolution of the population $P_{e}\left( t\right)=
|\left\langle 0\right\vert \hat{\sigma}_{-}\left\vert \Psi\left( t\right) \right\rangle|^2$ on the excited
2LE for different values of the coupling strengths.
When coupling strengths are much smaller than the hopping strength, the probability $P_{e}\left( t\right)$
exponentially decays as time increases, so the entire amount of the initial excitation in the 2LE is released
to the field and is propagated away to the left and right ends of the CRA. When coupling and hopping strengths
are of comparable values, secondary oscillations can be observed at intermediate time as shown in Fig.~\ref{fig5}(b)
and \ref{fig5}(d). At sufficiently long times, the time dependent $P_{e}\left( t\right)$ becomes a non-zero
constant in Fig.\ref{fig5}(b) and exhibits a stationary oscillation in Fig.\ref{fig5}(d), which demonstrates
that some amount of the excitation has been trapped within the 2LE owing to the presence of the bound states.
When coupling strengths are lager than the hopping strength, a persistent oscillation of population can be
observed, and the amplitude of this oscillation increases as shown in Fig.~\ref{fig5}(c). Actually, such a
persistent oscillation is also a stationary oscillation. The behavior of the 2LE's excitation can be understood
from Eq.~(\ref{4-02}). When $g_i/J\ll 1$, the contribution of the bound states is negligible as shown in Fig.~\ref{fig6},
so the energy is carried away from the 2LE by the scattering eigenstates. When $g_i$ approaches $J$, the
contribution of the bound states increases, but the number of the bound states determines the behavior of the
2LE at long enough time, i.e., the probability would be a constant if only one bound state is present as
shown in Fig.~\ref{fig5}(b), however, excitation trapped by two bound states leads to a residual oscillatory
behavior of the excited state population at long times as shown in Fig.~\ref{fig5}(d). When $g_i/J\gg 1$,
the scattering eigenstates contribution becomes negligible, the probability amplitude oscillates with
frequency $|E_+-E_-|/2\pi$.

\section{\label{Sec:5} Conclusion}
We consider a one-dimensional (1D) array of coupled resonators, where a 2LE is electric-dipole coupled to
the modes of two adjacent resonators. The energy spectrum in the single-excitation subspace is studied.
It consists of a finite band of energies, and out-of-band discrete levels. Different from the local coupling
between the 2LE and the 1D CRA where there are always two bound states, the number of the
discrete levels is controlled by the coupling strength and the transition frequency of the 2LE. The
change of energy spectrum reveals that there is a quantum phase transition. In addition, the wave functions
of the bound states are asymmetry around the 2LE when two coupling strengths have different magnitudes,
their preferred directions are determined by the larger one among the coupling strengths. The time evolution
of the 2LE's excitation is highlighted by three regimes. For every weak values of the coupling strengths,
the population of the excitation decays exponentially. At intermediate values of the coupling strengths,
the population of the excitation is trapped in the form of residual oscillation or residual non-zero
constant in the long enough time. When the coupling strengths are strong, the population of the excitation
persistently oscillates with larger amplitude. And the population trapping of the excitation is manifested
in the form of stationary oscillation or non-vanishing constant in the long enough time. When the dissipation is considered due to the system interacting with the environment, energies go out of the system, so the bound states has their lifetime, so they become quasi bound states.

\begin{acknowledgments}
NSFC Grants No. 11975095, No. 12075082, No.11935006, No. 12175150,,
the science and technology innovation Program of Hunan Province (Grant No. 2020RC4047) and Hunan Provincial major Sci-Tech Program(2023ZJ1010).
\end{acknowledgments}


\begin{thebibliography}{99}

\bibitem{BoydPRA69} D. D. Smith, H. Chang, K. A. Fuller,\emph{et al.}, Phys. Rev. A \textbf{69}, 063804 (2004).
\bibitem{KimPRA78} C. D. Ogden, E. K. Irish, and M. S. Kim, Phys. Rev. A \textbf{78}, 063805 (2008)

\bibitem{XPRL102} X. Yang, M. Yu, D.-L. Kwong, and Chee Wei Wong, \emph{et al.}, Phys. Rev. Lett. \textbf{102}, 173902 (2009).

\bibitem{Zhou08} L. Zhou, Z. R. Gong, Y.-x. Liu, \emph{et al.},
 Phys. Rev. Lett. \textbf{101}, 100501 (2008).

\bibitem{zhouPRA78(08)} L. Zhou, H. Dong, Y.-x. Liu,\emph{et al.},
 Phys. Rev. A \textbf{78}, 063827 (2008);

\bibitem{gongPRA78(08)} Z. R. Gong, H. Ian, L. Zhou, \emph{et al.},
 Phys. Rev. A \textbf{78}, 053806
(2008).

\bibitem{Zhou09} L. Zhou, S. Yang, Y.-x. Liu, \emph{et al.},
 Phys. Rev. A \textbf{80},
062109 (2009).

\bibitem{104(10)023602} P. Longo, P. Schmittechert, K. Busch,
 Rev. Lett. \textbf{104}, 023602 (2010).

\bibitem{zhouPRA85(12)} L. Zhou, Y. Chang, H. Dong, \emph{et al.},  Phys. Rev. A \textbf{85}, 013806 (2012).

\bibitem{Zhou13} L. Zhou, L. P. Yang, Y. Li, \emph{et al.}, Phys. Rev. Lett. \textbf{%
111}, 103604 (2013).

\bibitem{lujing89} J. Lu, L. Zhou, L.-M. Kuang, \emph{et al.},  Phys. Rev. A \textbf{89}, 013805 (2014).

\bibitem{wangPRA89} Z. H. Wang, L. Zhou, Y. Li , \emph{et al.},  Phys.
Rev. A \textbf{89}, 053813 (2014).

\bibitem{xuPRA17} X.-W. Xu, A.-X. Chen, Y. Li, \emph{et al.},  Phys. Rev. A \textbf{95}, 063808 (2017).

\bibitem{MahPRL123} S. Mahmoodian,  Phys. Rev. Lett. \textbf{123}, 133603 (2019).

\bibitem{LJinPRR4} H. S. Xu and L. Jin, Phys. Rev. Res. \textbf{4}, L032015 (2022).

\bibitem{PRX12BS} M. Scigliuzzo, G. Calaj, F. Ciccarello, \emph{et al.}, Phys. Rev. X \textbf{12}, 031036 (2022).

\bibitem{PRA102BS} E. V. Stolyarov, Phys. Rev. A \textbf{102}, 063709 (2020)

 \bibitem{PRA108 053717} R. Bag and D. Roy, Phys. Rev. A \textbf{108} ,053717(2023).

\bibitem{PRA108 013704} X. Zhang , \emph{et al.}, Phys. Rev. A  \textbf{108},013704(2023).

\bibitem{WangPRA101} W. Zhao and Z. H. Wang,  Phys. Rev. A \textbf{101}, 053855 (2020).

\bibitem{PalmaPRA89} F. Lombardo, F. Ciccarello, and G. M. Palma,  Phys.
Rev. A \textbf{89}, 053826 (2014).

\bibitem{PRA100(19)063806} L. Qiao and C.-P. Sun,  Phys. Rev. A \textbf{100}, 063806 (2019).

\bibitem{PRA96(17)023831} E. S\'{a}nchez-Burillo, D. Zueco, L. Mart\'{\i}%
n-Moreno, \emph{et al.},  Phys. Rev. A \textbf{96}, 023831 (2017).

\bibitem{BookQPT} S. Sachdev, Quantum Phase Transitions (Cambridge University
Press, Cambridge, UK, 1999).


\bibitem{NoriPRL126} X. Wang, T. Liu, A. F. Kockum, \emph{et al.},
 Phys. Rev. Lett. \textbf{126}, 043602 (2021).


\end{thebibliography}
\end{document}